# Exact Solutions of Nonlinear Schrodinger's Equation with Dual Power-Law Nonlinearity by Extended Trial Equation Method

Hasan BULUT[1], Yusuf PANDIR[2] and Seyma TULUCE DEMIRAY[1]

[1] Department of Mathematics, Firat University, 23119, Elazig, Turkey

[2] Department of Mathematics, Bozok University, 66100, Yozgat, Turkey

E-mail: hbulut@firat.edu.tr

**Abstract**

In this paper, we acquire the soliton solutions of the nonlinear Schrodinger's equation with dual power-law nonlinearity. Primiraly, we use the extended trial equation method to find exact solutions of this equation. Then, we attain some exact solutions including soliton solutions, rational and elliptic function solutions of this equation by using the extended trial equation method.

**Keywords**: The extended trial equation method, soliton solutions, rational solutions, elliptic function solutions and nonlinear Schrodinger's equation.

1. Introduction

The examination of exact solutions of nonlinear evolution equations (NLEEs) have a very important place in the enquiry of some physical phenomena. The types of solutions of NLEEs, that are combined utilizing variety mathematical techniques, are very significant various sciences such as chemistry, technology of space, control engineering problems, physics, applied mathematics and computer engineering. In this paper, the nonlinear Schrodinger's equation (NLSE) that arises in the study of long-distance optical communications and all-optical ultra fast switching devices will be investigated. Some solutions for this equation can be found in [1-4]. Also, Wazwaz have investigated reliable analysis for the (1+1)-dimensional NLSE with power law nonlinearity [28].

In 1926, Erwin Schrodinger found a new equation which is called time independent Schrodinger equation. This equation have illuminated sufficiently atomic phenomena and dynamical centerpiece of quantum wave mechanics. We assume that Erwin Schrodinger generated his equation based on three major principles such as de Broglie's hypothesis of matter wave, the law of conservation of energy and classical plane wave equation.

NLSE has been indicated to manage the evolution of a wave packet in a weakly nonlinear and dispersive medium and has eventuated diverse fields such as nonlinear optics, water waves and plasma physics. Another implementation of this equation is in pattern formation, where it has been used to model some nonequilibrium pattern forming systems. Especially, this equation is now widely used in the optics field as a good model for optical pulse propagation in nonlinear fibers [7].

Most previous studies focused attention on Kerr law nonlinearity media (the refractive index varies linearly with the beam intensity) [3]. Then, some of sciencist have been considered non-Kerr nonlinearity. One such non-Kerr law nonlinearity will be worked that is a generalization of the parabolic law nonlinearity which is called the dual-power law nonlinearity [5-7]. This type of non-Kerr nonlinerity have also been studied by A. Biswas and D. Milovic [2].

In recent years, most authors have improved a lot of methods to find exact solutions of NLEEs such as G'/G-expansion method [8-10], exp-function method [11-12], the tanh method [13], homogeneous balance method [14], and many more. In this paper, the extended trial equation method [15-27] will be applied to obtain exact solutions to the NLSE. Finally, we can say that the obtained solutions satisfy the equation.

**2. Fundamentals of the Extended Trial Equation Method**

**Step 1.** For a known nonlinear partial differential equation

$$P(u, u_t, u_x, u_{xx}, \ldots) = 0 \tag{1}$$

get the wave transformation

$$u(x_1, x_2, \ldots, x_N, t) = u(\eta), \quad \eta = \lambda\left(\sum_{j=1}^{N} x_j - ct\right), \tag{2}$$

where $\lambda \neq 0$ and $c \neq 0$. Putting Eq. (2) into Eq. (1) satisfies a nonlinear ordinary differential equation,

$$N(u, u', u'', \ldots) = 0. \tag{3}$$

**Step 2.** Take transformation and trial equation as following:

$$u = \sum_{i=0}^{\delta} \tau_i \Gamma^i, \tag{4}$$

where

$$(\Gamma')^2 = \Lambda(\Gamma) = \frac{\phi(\Gamma)}{\psi(\Gamma)} = \frac{\xi_\theta \Gamma^\theta + \ldots + \xi_1 \Gamma + \xi_0}{\zeta_\epsilon \Gamma^\epsilon + \ldots + \zeta_1 \Gamma + \zeta_0}. \tag{5}$$

Taking into consideration relations (4) and (5), we can get

$$(u')^2 = \frac{\phi(\Gamma)}{\psi(\Gamma)}\left(\sum_{i=0}^{\delta} i\tau_i \Gamma^{i-1}\right)^2, \tag{6}$$

$$u'' = \frac{\phi'(\Gamma)\psi(\Gamma) - \phi(\Gamma)\psi'(\Gamma)}{2\psi^2(\Gamma)}\left(\sum_{i=0}^{\delta} i\tau_i \Gamma^{i-1}\right) + \frac{\phi(\Gamma)}{\psi(\Gamma)}\left(\sum_{i=0}^{\delta} i(i-1)\tau_i \Gamma^{i-2}\right) \tag{7}$$

where $\phi(\Gamma)$ and $\psi(\Gamma)$ are polynomials. Embedding these terms into Eq. (3) provides an equation of polynomial $\Omega(\Gamma)$ of $\Gamma$:

$$\Omega(\Gamma) = \sigma_s \Gamma^s + \ldots + \sigma_1 \Gamma + \sigma_0 = 0. \tag{8}$$

With respect to balance principle, we can define a relation of $\theta, \in$ and $\delta$. We can get some values of $\theta, \in$ and $\delta$.

**Step 3.** Let the coefficients of $\Omega(\Gamma)$ all be zero will satisfy an algebraic equations system:

$$\sigma_i = 0, \quad i = 0, \ldots s. \tag{9}$$

Solving this equation system (9), we will define the values of $\xi_0, \ldots, \xi_\theta$; $\zeta_0, \ldots, \zeta_\in$ and $\tau_0, \ldots, \tau_\delta$.

**Step 4.** Simplify Eq. (5) to elementary integral shape,

$$\pm(\eta - \eta_0) = \int \frac{d\Gamma}{\sqrt{\Lambda(\Gamma)}} = \int \sqrt{\frac{\psi(\Gamma)}{\phi(\Gamma)}} d\Gamma. \tag{10}$$

Applying a complete discrimination system for polynomial to classify the roots of $\phi(\Gamma)$, we solve the infinite integral (10) and categorize the exact solutions to Eq. (1) by Mathematica [18].

### 3. Implementation of the Extended Trial Equation Method to Nonlinear Schrodinger's Equation

We consider the following NLSE equation[1-4],

$$iq_t + \frac{1}{2}(q_{xx} + q_{yy}) + \left(|q|^{2m} + k|q|^{4m}\right)q = 0, \tag{11}$$

where $q(x, y, t)$ is a complex function that indicates the complex amplitude of the wave form, m is an arbitrary nonzero constant, k is a constant which indicates the saturation of the nonlinear refractive index.

In an attempt to find travelling wave solutions of the Eq. (11), we take the transformation using the wave variables

$$q(x, y, t) = e^{i\chi} u(\omega), \ \chi = \chi_1 x + \chi_2 y + \chi_3 t, \ \omega = \omega_1 x + \omega_2 y + \omega_3 t, \tag{12}$$

where $\chi_1$, $\chi_2$, $\chi_3$, $\omega_1$, $\omega_2$, $\omega_3$ are arbitrary constants. $\omega_1$ and $\omega_2$ are the inverse width in the x and y directions, respectively. $\omega_3$ represents the velocity of the soliton. Moreover, $\chi_1$ and $\chi_2$ represents the soliton frequency in the x and y directions, respectively, while $\chi_3$ represents the solitary wave number.

Substituting the following (13-15) derivatives into Eq.(11),

$$iq_t = ie^{i\chi}\omega_3 u' - e^{i\chi}\chi_3 u, \tag{13}$$

$$q_{xx} = -\chi_1^2 e^{i\chi} u + 2ie^{i\chi}\chi_1\omega_1 u' + \omega_1^2 e^{i\chi} u'', \tag{14}$$

$$q_{yy} = -\chi_2^2 e^{i\chi} u + 2ie^{i\chi}\chi_2\omega_2 u' + \omega_2^2 e^{i\chi} u'', \tag{15}$$

we get

$$ie^{i\chi}\omega_3 u' - e^{i\chi}\chi_3 u + \frac{e^{i\chi}}{2}\left(2i\chi_1\omega_1 u' - \chi_1^2 u + +\omega_1^2 u'' + 2i\chi_2\omega_2 u' - \chi_2^2 u + \omega_2^2 u''\right) + e^{i\chi}\left(u^{2m+1} + ku^{4m+1}\right) = 0. \tag{16}$$

Then we obtain following system from Eq.(16),

$$i(\omega_3 + \chi_1\omega_1 + \chi_2\omega_2)u'(\omega) = 0, \tag{17}$$

$$\left(-\chi_3 - \frac{\chi_1^2}{2} - \frac{\chi_2^2}{2}\right)u + \left(\frac{\omega_1^2}{2} + \frac{\omega_2^2}{2}\right)u'' + u^{2m+1} + ku^{4m+1} = 0. \tag{18}$$

Taking into consideration the transformation

$$u = v^{\frac{1}{2m}}, \tag{19}$$

Eq. (18) converts into the equation

$$\left(-2\chi_3 - \chi_1^2 - \chi_2^2\right)4m^2 v^2 + \left(\omega_1^2 + \omega_2^2\right)(1-2m)(v')^2 + \left(\omega_1^2 + \omega_2^2\right)2mvv'' + 8m^2 v^3 + 8km^2 v^4 = 0 \tag{20}$$

Embedding Eqs. (6) and (7) into Eq. (20), and using the balance principle, we obtain

$$\theta = 2\delta + \epsilon + 2. \tag{21}$$

In attempt to obtain the exact solution of Eq.(11), if we choose $\epsilon = 0, \delta = 1$ and $\theta = 4$ in Eq.(21), then

$$(v')^2 = \frac{\tau_1^2\left(\xi_0 + \Gamma\xi_1 + \Gamma^2\xi_2 + \Gamma^3\xi_3 + \Gamma^4\xi_4\right)}{\zeta_0}, \quad v'' = \frac{\tau_1\left(\xi_1 + 2\Gamma\xi_2 + 3\Gamma^2\xi_3 + 4\Gamma^3\xi_4\right)}{2\zeta_0}, \tag{22}$$

where $\xi_4 \neq 0, \zeta_0 \neq 0$. Respectively, solving the algebraic equation system (9) satisfies

$$\xi_0 = -\frac{k(1+m)\xi_3\tau_0\left(k^2(1+m)^2\xi_3^3\tau_0^2(1+2m+2k(1+m)\tau_0) - \xi_1\xi_4^2\Upsilon^3\right)}{2\xi_4^3\Upsilon^4},$$

$$\xi_1 = \xi_1, \quad \xi_2 = \frac{\xi_1\xi_4^2\Upsilon^3 + k^2(1+m)^2\xi_3^3\tau_0^2(3+6m+8k(1+m)\tau_0)}{2k(1+m)\xi_3\xi_4\tau_0\Upsilon^2}, \tag{23}$$

$$\xi_3 = \xi_3, \quad \xi_4 = \xi_4, \quad \zeta_0 = \frac{k(1+m)^2(1+2m)(\omega_1^2 + \omega_2^2)\xi_3^2}{8m^2\xi_4\Upsilon^2}, \quad \tau_1 = \frac{\xi_4\Upsilon}{k(1+m)\xi_3}, \quad \tau_0 = \tau_0,$$

$$\theta_3 = -\frac{\xi_1 \xi_4^2 \Upsilon^3 + k^2(1+m)^2 \xi_3^3 \tau_0 \left((1+m)(1+2m)\left(\chi_1^2 + \chi_2^2\right) - 3(1+2m)\tau_0 - 4k(1+m)\tau_0^2\right)}{2k^2(1+m)^3(1+2m)\xi_3^3 \tau_0},$$

where $\Upsilon = (1+2m+4k(1+m)\tau_0)$.

Substituting these results into Eqs. (5) and (10), we get

$$\pm(\eta - \eta_0) = A\int \frac{d\Gamma}{\sqrt{\frac{\xi_0}{\xi_4} + \frac{\xi_1}{\xi_4}\Gamma + \frac{\xi_2}{\xi_4}\Gamma^2 + \frac{\xi_3}{\xi_4}\Gamma^3 + \Gamma^4}} \tag{24}$$

where $A = \sqrt{\dfrac{k(1+m)^2(1+2m)\left(\omega_1^2 + \omega_2^2\right)\xi_3^2}{8m^2 \xi_4^2 \Upsilon^2}}.$

Integrating Eq. (24), we obtain the solutions of Eq. (11) as following:

$$\pm(\eta - \eta_0) = -\frac{A}{\Gamma - \alpha_1}, \tag{25}$$

$$\pm(\eta - \eta_0) = \frac{2A}{\alpha_1 - \alpha_2}\sqrt{\frac{\Gamma - \alpha_2}{\Gamma - \alpha_1}}, \quad \alpha_1 > \alpha_2, \tag{26}$$

$$\pm(\eta - \eta_0) = \frac{A}{\alpha_1 - \alpha_2}\ln\left|\frac{\Gamma - \alpha_1}{\Gamma - \alpha_2}\right|, \quad \alpha_1 > \alpha_2, \tag{27}$$

$$\pm(\eta - \eta_0) = \frac{2A}{\sqrt{(\alpha_1 - \alpha_2)(\alpha_1 - \alpha_3)}}\ln\left|\frac{\sqrt{(\Gamma - \alpha_2)(\alpha_1 - \alpha_3)} - \sqrt{(\Gamma - \alpha_3)(\alpha_1 - \alpha_2)}}{\sqrt{(\Gamma - \alpha_2)(\alpha_1 - \alpha_3)} + \sqrt{(\Gamma - \alpha_3)(\alpha_1 - \alpha_2)}}\right|, \quad \alpha_1 > \alpha_2 > \alpha_3, \tag{28}$$

$$\pm(\eta - \eta_0) = \frac{2A}{\sqrt{(\alpha_1 - \alpha_3)(\alpha_2 - \alpha_4)}} F(\varphi, l), \quad \alpha_1 > \alpha_2 > \alpha_3 > \alpha_4, \tag{29}$$

where

$$F(\varphi, l) = \int_0^\varphi \frac{d\psi}{\sqrt{1 - l^2 \sin^2 \psi}}, \quad \varphi = \arcsin\sqrt{\frac{(\Gamma - \alpha_1)(\alpha_2 - \alpha_4)}{(\Gamma - \alpha_2)(\alpha_1 - \alpha_4)}}, \quad l^2 = \frac{(\alpha_2 - \alpha_3)(\alpha_1 - \alpha_4)}{(\alpha_1 - \alpha_3)(\alpha_2 - \alpha_4)}. \tag{30}$$

Also, $\alpha_1, \alpha_2, \alpha_3$ and $\alpha_4$ are the roots of the polynomial equation

$$\Gamma^4 + \frac{\xi_3}{\xi_4}\Gamma^3 + \frac{\xi_2}{\xi_4}\Gamma^2 + \frac{\xi_1}{\xi_4}\Gamma + \frac{\xi_0}{\xi_4} = 0. \tag{31}$$

Substituting the solutions (25-29) into (4) and (19), we obtain $u$ functions in terms of $\Gamma$ the following, respectively:

$$u_1(\omega) = \left[ \tau_0 + \tau_1 \alpha_1 \pm \frac{\tau_1 A}{\omega_1 x + \omega_2 y + \omega_3 t - \eta_0} \right]^{\frac{1}{2m}}, \tag{32}$$

$$u_2(\omega) = \left[ \tau_0 + \tau_1 \alpha_1 + \frac{4A^2 (\alpha_2 - \alpha_1) \tau_1}{4A^2 - \left[ (\alpha_1 - \alpha_2)(\omega_1 x + \omega_2 y + \omega_3 t - \eta_0) \right]^2} \right]^{\frac{1}{2m}}, \tag{33}$$

$$u_3(\omega) = \left[ \tau_0 + \tau_1 \alpha_2 + \frac{(\alpha_2 - \alpha_1) \tau_1}{\exp\left[ \frac{(\alpha_1 - \alpha_2)}{A} (\omega_1 x + \omega_2 y + \omega_3 t - \eta_0) \right] - 1} \right]^{\frac{1}{2m}}, \tag{34}$$

$$u_4(\omega) = \left[ \tau_0 + \tau_1 \alpha_1 + \frac{(\alpha_1 - \alpha_2) \tau_1}{\exp\left[ \frac{(\alpha_1 - \alpha_2)}{A} (\omega_1 x + \omega_2 y + \omega_3 t - \eta_0) \right] - 1} \right]^{\frac{1}{2m}}, \tag{35}$$

$$u_5(\omega) = \left[ \tau_0 + \tau_1 \alpha_1 - \frac{2(\alpha_1 - \alpha_2)(\alpha_1 - \alpha_3) \tau_1}{2\alpha_1 - \alpha_2 - \alpha_3 + (\alpha_3 - \alpha_2) \cosh\left[ \frac{\sqrt{(\alpha_1 - \alpha_2)(\alpha_1 - \alpha_3)}}{A} (\omega_1 x + \omega_2 y + \omega_3 t - \eta_0) \right]} \right]^{\frac{1}{2m}}, \tag{36}$$

$$u_6(\omega) = \left[ \tau_0 + \tau_1 \alpha_2 + \frac{2(\alpha_1 - \alpha_2)(\alpha_1 - \alpha_3) \tau_1}{\alpha_4 - \alpha_2 + (\alpha_1 - \alpha_4) sn^2 \left[ \mp \frac{\sqrt{(\alpha_1 - \alpha_3)(\alpha_2 - \alpha_4)}}{2A} (\omega_1 x + \omega_2 y + \omega_3 t - \eta_0), \frac{(\alpha_2 - \alpha_3)(\alpha_1 - \alpha_4)}{(\alpha_1 - \alpha_3)(\alpha_2 - \alpha_4)} \right]} \right]^{\frac{1}{2m}}, \tag{37}$$

If we take $\tau_0 = -\tau_1 \alpha_1$, and $\eta_0 = 0$, then the above solutions (32)-(33) and (35)-(36) can reduce to more simple circumstance

$$u_1(\omega) = \left( \pm \frac{A_1}{\omega_1 x + \omega_2 y + \omega_3 t} \right)^{\frac{1}{2m}}, \tag{38}$$

$$u_2(\omega) = \left( \frac{4A^2 (\alpha_2 - \alpha_1) \tau_1}{4A^2 - \left[ (\alpha_1 - \alpha_2)(\omega_1 x + \omega_2 y + \omega_3 t) \right]^2} \right)^{\frac{1}{2m}}, \tag{39}$$

$$u_3(\omega) = \left(\frac{(\alpha_2 - \alpha_1)\tau_1}{2}\left\{1 \pm \coth\left[\frac{(\alpha_1 - \alpha_2)}{A}(\omega_1 x + \omega_2 y + \omega_3 t)\right]\right\}\right)^{\frac{1}{2m}}, \quad (40)$$

$$u_4(\omega) = \frac{A_2}{\left(D + \cosh\left[B(\omega_1 x + \omega_2 y + \omega_3 t)\right]\right)^{\frac{1}{2m}}}, \quad (41)$$

where $A_1 = \tau_1 A$, $A_2 = \left(\frac{2\tau_1(\alpha_1 - \alpha_2)(\alpha_1 - \alpha_3)}{\alpha_3 - \alpha_2}\right)^{\frac{1}{2m}}$, $B = \frac{\sqrt{(\alpha_1 - \alpha_2)(\alpha_1 - \alpha_3)}}{A}$,

$D = \frac{2\alpha_1 - \alpha_2 - \alpha_3}{\alpha_3 - \alpha_2}$. Here, $A_2$ is the amplitude of the soliton, and $B$ is the inverse width of the solitons. Thus, we can say that the solitons exist for $\tau_1 < 0$.

On the other hand, if we take $\tau_0 = -\tau_1 \alpha_2$, and $\eta_0 = 0$, the Jacobi elliptic function solution (37) can be written in the form

$$u_5(\omega) = \frac{A_3}{\left(M + N\, sn^2(\varphi, l)\right)^{\frac{1}{2m}}}, \quad (42)$$

where $A_3 = \left(2\tau_1(\alpha_1 - \alpha_2)(\alpha_1 - \alpha_3)\right)^{\frac{1}{2m}}$, $M = \alpha_4 - \alpha_2$, $N = \alpha_1 - \alpha_4$, $l^2 = \frac{(\alpha_2 - \alpha_3)(\alpha_1 - \alpha_4)}{(\alpha_1 - \alpha_3)(\alpha_2 - \alpha_4)}$,

$$\varphi = \frac{\pm\sqrt{(\alpha_1 - \alpha_3)(\alpha_2 - \alpha_4)}}{2A}(\omega_1 x + \omega_2 y + \omega_3 t).$$

As a consequently, the solutions of Eq. (11) are obtained by using Eq. (12) as following,

$$q_1(x, y, t) = e^{i(\chi_1 x + \chi_2 y + \chi_3 t)}\left(\pm \frac{A_1}{\omega_1 x + \omega_2 y + \omega_3 t}\right)^{\frac{1}{2m}}, \quad (43)$$

$$q_2(x, y, t) = e^{i(\chi_1 x + \chi_2 y + \chi_3 t)}\left(\frac{4A^2(\alpha_2 - \alpha_1)\tau_1}{4A^2 - \left[(\alpha_1 - \alpha_2)(\omega_1 x + \omega_2 y + \omega_3 t)\right]^2}\right)^{\frac{1}{2m}}, \quad (44)$$

$$q_3(x, y, t) = e^{i(\chi_1 x + \chi_2 y + \chi_3 t)}\left(\frac{(\alpha_2 - \alpha_1)\tau_1}{2}\left\{1 \pm \coth\left[\frac{(\alpha_1 - \alpha_2)}{A}(\omega_1 x + \omega_2 y + \omega_3 t)\right]\right\}\right)^{\frac{1}{2m}}, \quad (45)$$

$$q_4(x,y,t) = e^{i(\chi_1 x + \chi_2 y + \chi_3 t)} \frac{A_2}{\left(D + \cosh\left[B\left(\omega_1 x + \omega_2 y + \omega_3 t\right)\right]\right)^{\frac{1}{2m}}}, \qquad (46)$$

$$q_5(x,y,t) = e^{i(\chi_1 x + \chi_2 y + \chi_3 t)} \frac{A_3}{\left(M + N\,sn^2(\varphi, l)\right)^{\frac{1}{2m}}}. \qquad (47)$$

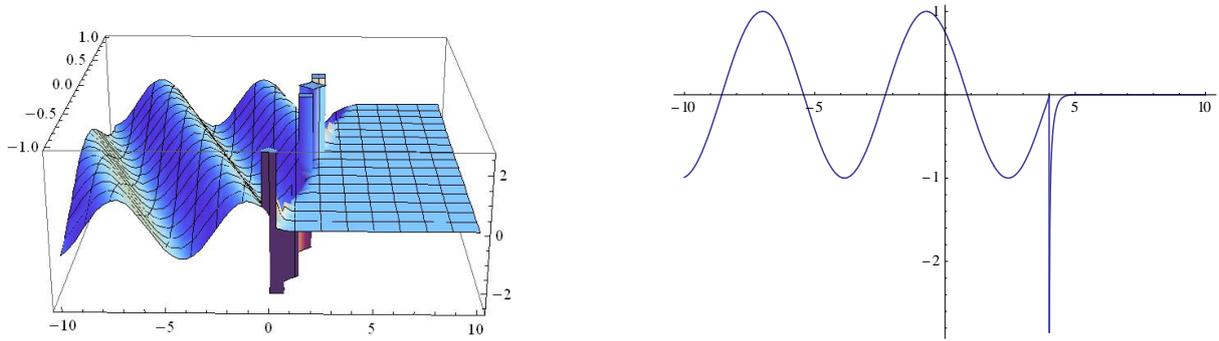

Fig.1. Graph of imaginer values of the solution (45) is shown at $m = k = \tau_0 = \tau_1 = \xi_3 = \xi_4 = \alpha_2 = \omega_2 = \chi_1 = 1$, $\alpha_1 = \omega_3 = \chi_2 = 2, \chi_3 = 3, \omega_1 = -1$ and the second graph represents imaginer values of exact solution of Eq. (45) for $t = 1$.

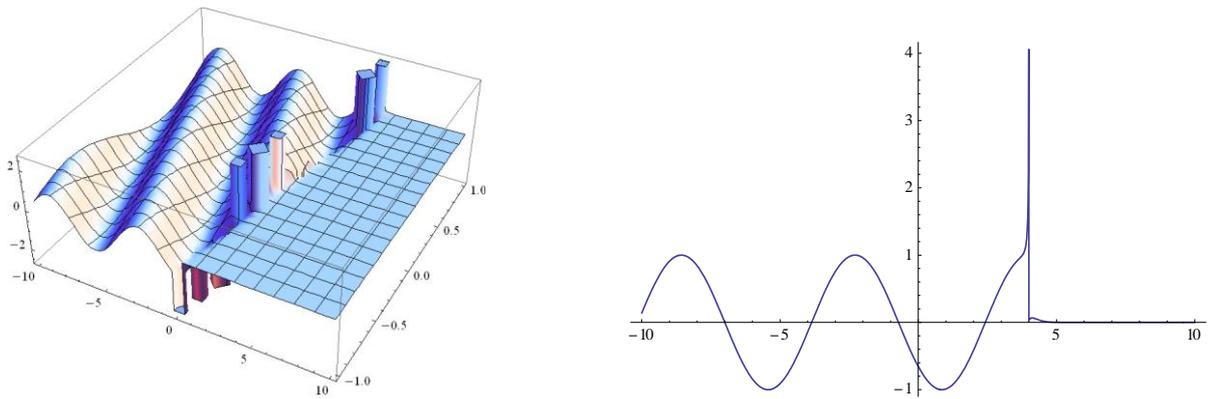

Fig.2. Graph of real values of the solution (45) is shown at $m = k = \tau_0 = \tau_1 = \xi_3 = \xi_4 = \alpha_2 = \omega_2 = \chi_1 = 1$, $\alpha_1 = \omega_3 = \chi_2 = 2, \chi_3 = 3, \omega_1 = -1$ and the second graph represents real values of exact solution of Eq. (45) for $t = 1$.

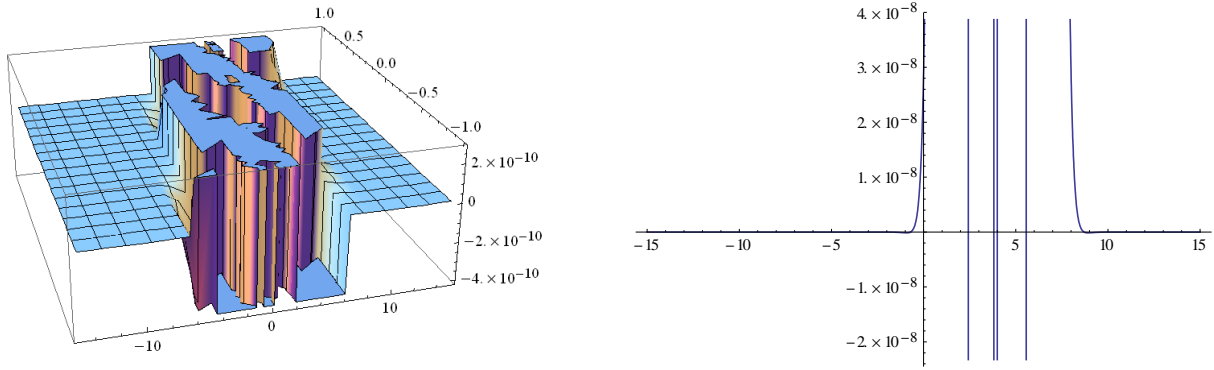

Fig.3. Graph of imaginer values of the solution (46) is shown at $m = k = \tau_0 = \tau_1 = \xi_3 = \xi_4 = \alpha_1 = \omega_2 = \chi_1 = 1$, $\alpha_2 = \omega_3 = \chi_2 = 2, \chi_3 = 3, \omega_1 = -1, \alpha_3 = 3$ and the second graph represents imaginer values of exact solution of Eq. (46) for $t = 1$.

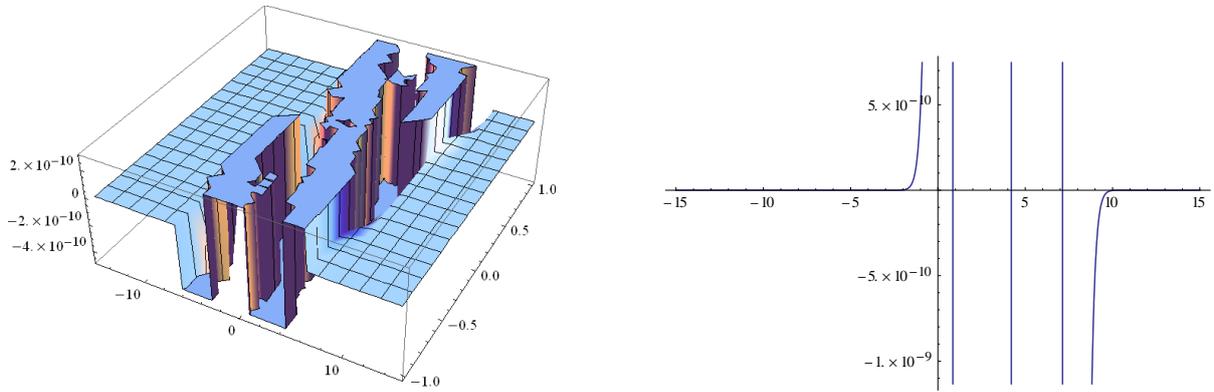

Fig.4. Graph of real values of the solution (46) is shown at $m = k = \tau_0 = \tau_1 = \xi_3 = \xi_4 = \alpha_1 = \omega_2 = \chi_1 = 1$, $\alpha_2 = \omega_3 = \chi_2 = 2, \chi_3 = 3, \omega_1 = -1, \alpha_3 = 3$ and the second graph represents real values of exact solution of Eq. (46) for $t = 1$.

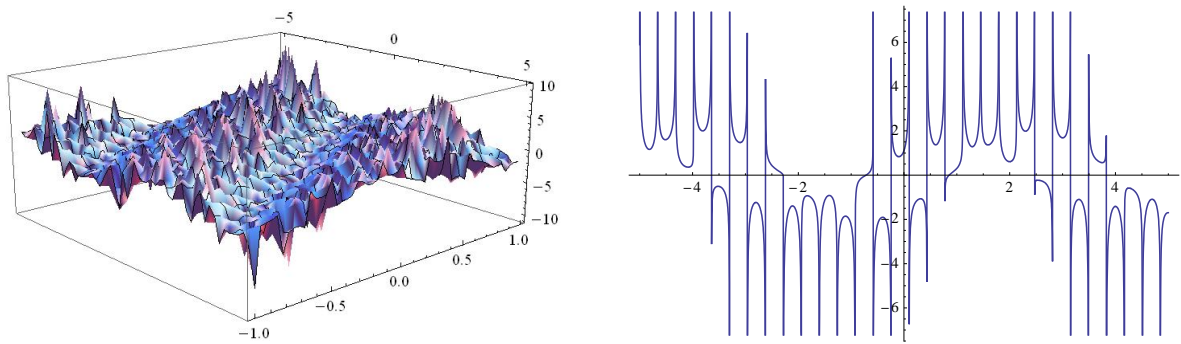

Fig.5. Graph of imaginer values of the solution (47) is shown at $m = k = \tau_0 = \tau_1 = \xi_3 = \xi_4 = \alpha_1 = \omega_2 = \chi_1 = 1$, $\alpha_2 = \omega_3 = \chi_2 = 2, \chi_3 = 3, \omega_1 = -1, \alpha_3 = 3, \alpha_4 = 4$ and the second graph represents imaginer values of exact solution of Eq. (47) for $t = 1$.

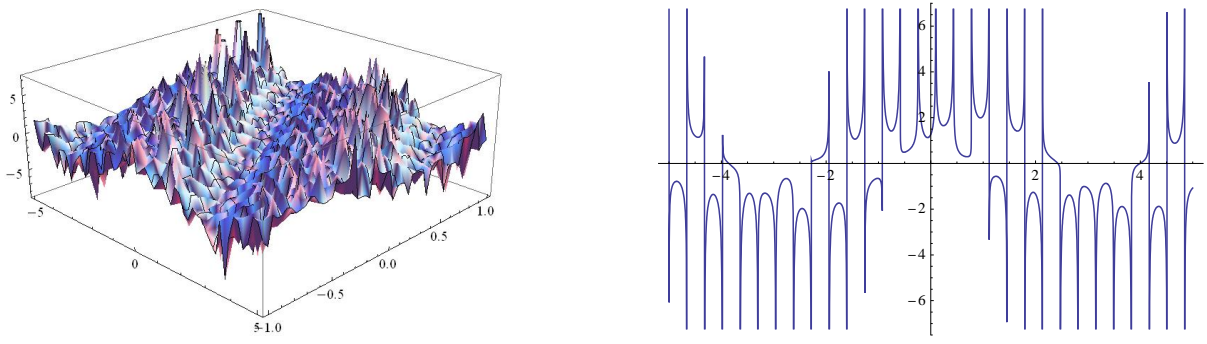

Fig.6. Graph of real values of the solution (47) is shown at $m = k = \tau_0 = \tau_1 = \xi_3 = \xi_4 = \alpha_1 = \omega_2 = \chi_1 = 1$, $\alpha_2 = \omega_3 = \chi_2 = 2, \chi_3 = 3, \omega_1 = -1, \alpha_3 = 3, \alpha_4 = 4$ and the second graph represents real values of exact solution of Eq. (47) for $t = 1$.

**Remark 1.** When the modulus $l \to 1$, then the solution (42) can be converted into the hyperbolic function solution of the nonlinear Schrodinger's equation.

$$u_6(\omega) = \frac{A_3}{\left(M + N \tanh^2\left[\frac{\sqrt{(\alpha_1 - \alpha_3)(\alpha_2 - \alpha_4)}}{2A}(\omega_1 x + \omega_2 y + \omega_3 t)\right]\right)^{\frac{1}{2m}}}, \quad (48)$$

where $\alpha_3 = \alpha_4$. By using Eq. (12), we get

$$q_6(x, y, t) = e^{i(\chi_1 x + \chi_2 y + \chi_3 t)} \frac{A_3}{\left(M + N \tanh^2\left[\frac{\sqrt{(\alpha_1 - \alpha_3)(\alpha_2 - \alpha_4)}}{2A}(\omega_1 x + \omega_2 y + \omega_3 t)\right]\right)^{\frac{1}{2m}}}. \quad (49)$$

**Remark 2.** When the modulus $l \to 0$, then the solution (42) can be reduced to the periodic wave solution

$$u_7(\omega) = \frac{A_3}{\left(M + N \sin^2\left[\frac{\sqrt{(\alpha_1 - \alpha_3)(\alpha_2 - \alpha_4)}}{2A}(\omega_1 x + \omega_2 y + \omega_3 t)\right]\right)^{\frac{1}{2m}}}, \quad (50)$$

where $\alpha_2 = \alpha_3$. By using Eq. (12), we obtain

$$q_7(x, y, t) = e^{i(\chi_1 x + \chi_2 y + \chi_3 t)} \frac{A_3}{\left(M + N \sin^2\left[\frac{\sqrt{(\alpha_1 - \alpha_3)(\alpha_2 - \alpha_4)}}{2A}(\omega_1 x + \omega_2 y + \omega_3 t)\right]\right)^{\frac{1}{2m}}}. \quad (51)$$

**Remark 3.** We have obtained the similar solutions with the solution equation (16) existing in "1-soliton solution of (1+2) –dimensional nonlinear Schrödinger's equation in dual-power law media" written by Anjan Biswas and the solution equation (34) existing in "Optical solitons in 1+2 dimensions with non-Kerr law nonlinearity" written by A. Biswas and D. Milovic in this study as $q_4$ within solution equation (46). Moreover, we have also presented the solution equations of $q_1, q_2, q_3, q_5, q_6, q_7$ that do not exist in literature.

### 4. Conclusion

You can see that we found exact solutions of NLSE with dual power-law nonlinearity. This dual power-law nonlinearity model is utilized to define the saturation of the nonlinear refractive index, and its exact soliton solutions are known. The effective NLSE with this dual-power law nonlinearity serves as a basic model to identify spatial solitons in photovoltaic-photorefractive materials such as lithium niobate. Optical nonlinearities in many organic and polymer materials can be modelled using this form of nonlinearity [7].

In this paper, we obtained the solutions of Jacobi elliptic function, hyperbolic function and periodic wave of nonlinear Schrodinger's Equation by using extended trial equation method. By these solutions, we contributed new solutions that are obtained as $q_1, q_2, q_3, q_5, q_6, q_7$ by comparing existed solutions in literature. Three and two dimensional graphics of these new solutions have been plotted. According to the obtained data, it has been seen that extended trial equation method has been effective for the analytical solutions of nonlinear Schrödinger's equation and it is also effective to find new solutions.

### References

[1] Biswas A 2008 1-soliton solution of (1+2)-dimensional nonlinear Schrödinger's equation in dual-power law media *Physics Letters A* **372** 5941–5943.
[2] Biswas A, Milovic D 2009 Optical solitons in 1+2 dimensions with non-Kerr law nonlinearity *Eur. Phys. J. Special Topics* **173** 81–86.
[3] Zhang Li-Hua, Si Jian-Guo 2010 New soliton and periodic solutions of (1+2)-dimensional nonlinear Schrödinger equation with dual-power law nonlinearity *Commun Nonlinear Sci Numer Simulat* **15** 2747–2754.
[4] Liu H, Yan F, Xu C 2012 The bifurcation and exact travelling wave solutions of *(1+2)*-dimensional nonlinear Schrödinger equation with dual-power law nonlinearity *Nonlinear Dyn* **67** 465–473.
[5] Akhmediev N, Ankiewicz A and Grimshaw R 1999 Hamiltonian versus energy diagrams in soliton theory *Phys. Rev. E* **59**(5) 6088 - 6096.

[6] Biswas A 2003 Quasi-Stationary non-Kerr law optical solitons, *Optical Fiber Technology* **9(4)** 224-259.

[7] Biswas A and Konar S 2006 *Introduction to non-Kerr law optical solitons*, CRC Press, Boca Raton, FL.

[8] Wang M, Li X, Zhang J 2008 The G'/G-expansion method and travelling wave solutions of nonlinear evolution equations in mathematical physics *Phys. Lett. A* **372** 417-423.

[9] Ebadi G, Biswas A 2011 The G'/G method and topological soliton solution of the K(m,n) equation *Commun. Nonlinear Sci. Numer. Simulat.* **16** 2377-2382.

[10] Gurefe Y, Misirli E 2011 New variable separation solutions of two-dimensional Burgers system *Appl. Math. Comput.* **217(22)** 9189-9197.

[11] He JH, Wu XH 2006 Exp-function method for nonlinear wave equations Chaos *Soliton. Fract.* **30** 700-708.

[12] Gurefe Y, Misirli E 2011 Exp-function method for solving nonlinear evolution equations with higher order nonlinearity *Comput. Math. Appl.* **61** 2025-2030.

[13] Fan E 2000 Extended tanh-function method and its applications to nonlinear equations *Phys. Lett. A* **277(4-5)** 212-218.

[14] Wang M 1995 Solitary wave solutions for variant Boussinesq equations *Phys. Lett. A* **199(3-4)** 169-172.

[15] Liu CS 2005 Trial equation method and its applications to nonlinear evolution equations *Acta Phys. Sin.-Ch. Ed.* **54(6)** 2505.

[16] Du XH 2010 An irrational trial equation method and its applications *Pramana-J. Phys.* **75(3)** 415-422.

[17] Gurefe Y, Sonmezoglu A, Misirli E 2011 Application of the trial equation method for solving some nonlinear evolution equations arising in mathematical physics *Pramana-J. Phys.* **77(6)** 1023-1029.

[18] Gurefe Y, Sonmezoglu A, Misirli E 2012 Application of an irrational trial equation method to high-dimensional nonlinear evolution equations *J. Adv. Math. Stud.* **5(1)** 41-47.

[19] Pandir Y, Gurefe Y, Kadak U, Misirli E 2012 Classifications of exact solutions for some nonlinear partial differential equations with generalized evolution *Abstr. Appl. Anal.* **2012** 1-16.

[20] Pandir Y, Gurefe Y, Misirli E 2013 Classification of exact solutions to the generalized Kadomtsev–Petviashvili equation *Physica Scripta* **87** 025003 (12pp).

[21] Pandir Y, Gurefe Y, Misirli E 2013 The extended trial equation method for some time fractional differential equations *Discrete Dynamics in Nature and Society* **2013** Article ID 491359, http://dx.doi.org/10.1155/2013/491359, 14 pages.

[22] Pandir Y, Gurefe Y 2013 New exact solutions of the generalized fractional Zakharov-Kuznetsov equations *Life Science Journal* **10 (2)** 2701–2705.

[23] Bulut H, Baskonus H. M, Pandir Y 2013 The modified trial equation method for fractional wave equation and time-fractional generalized Burgers equation *Abstract and Applied Analysis* **2013** Article ID 636802, 13 pp.

[24] Pandir Y, Gurefe Y, Misirli E 2013 A multiple extended trial equation method for the fractional Sharma-Tasso-Olver equation *AIP Conference Proceedings* **1558** 1927–1930 http://dx.doi.org/10.1063/1.4825910.

[25] Bulut H, Pandir Y 2013 Modified trial equation method to the nonlinear fractional Sharma–Tasso–Olever equation *International Journal of Modeling and Optimization* **3(4)** 353–357.

[26] Bulut H, Baskonus H. M and Pandir Y 2013 The Modified Trial Equation Method for Fractional Wave Equation and Time-Fractional Generalized Burgers Equation *Abstract and Applied Analysis* **2013** 1-8 pages.

[27] Bulut H 2013 Classification of exact solutions for generalized form of K(m,n) equation *Abstract and Applied Analysis* **2013** 1-11 pages.

[28] Wazwaz A M 2006 Reliable analysis for nonlinear Schrödinger equations with a cubic nonlinearity and power law nonlinearity *Math Comput Model* **43** 178-184.